**Main Belt Asteroid Science in the Decade 2023-2032: Fundamental Science Questions and Recommendations on behalf of the Small Bodies Assessment Group**

*Authors:*
Maggie M. McAdam (1), Andrew S. Rivkin (2), Lucy F. Lim (3), Julie Castillo-Rogez (4), Franck Marchis (5), Tracy M. Becker (6)

*Cosigners:*
Edgard G. Rivera-Valentín (7), Maitrayee Bose (8), Michelle Thompson (9), Conel M. O'D. Alexander (10), Jamie L. Molaro (11) Michael C. Nolan (12) Noam R. Izenberg (2), Henry H. Hsieh (11), Caitlin Ahrens (13), Jessica Noviello (8), Cristina Thomas (14), Hannah Kaplan (3), Michael S. Bramble (4), Meenakshi Wadhwa (8).

*Affiliations:*
1. NASA Ames Research Center
2. Johns Hopkins University Applied Physics Laboratory
3. NASA Goddard Space Flight Center
4. Jet Propulsion Laboratory, California Institute of Technology
5. SETI Institute
6. Southwest Research Institute
7. Lunar and Planetary Institute (USRA)
8. Arizona State University
9. Purdue University
10. EPL, Carnegie Institution of Washington
11. Planetary Science Institute
12. Lunar and Planetary Laboratory, University of Arizona
13. University of Arkansas
14. Northern Arizona University

*Lead Authors Contact Information:*
Maggie M. McAdam
Building N232
NASA Ames Research Center,
Moffett Field,CA 94035
Ph: 603-305-3783
maggie.mcadam@nasa.gov

Andrew. S Rivkin
JHU Applied Physics Laboratory
Ph: 240-228-2811
Andy.Rivkin@jhuapl.edu

*Executive Summary:* Solicited by the Small Bodies Assessment Group, we present and discuss the small body science community's fundamental questions and recommendations for Main Belt asteroid science in the upcoming decade. We briefly and non-exhaustively review some of the breakthroughs in our field during the last decade and discuss what important science still remains. The fundamental questions facing Main Belt asteroid science fall into three groups: physical properties and processes, chemical composition and evolution and dynamical evolution. We recommend a balanced program of telescopic observation (ground-based, airborne, and space-based), laboratory studies, theoretical research and missions to Main Belt Asteroids utilizing the full spectral range from ultraviolet to far-infrared to investigate these outstanding fundamental questions in the next decade.

## *Background and Motivation*

The decade of the 2010s saw great advancements in our understanding of solar system formation and evolution. Observations from telescopes and spacecraft combined with laboratory geochemical measurements of meteorites and collisional and dynamical modeling of individual objects and populations give us greater insight into the overall story of how our planetary system was put together, a story which features main-belt asteroids as key witnesses and objects for further study.

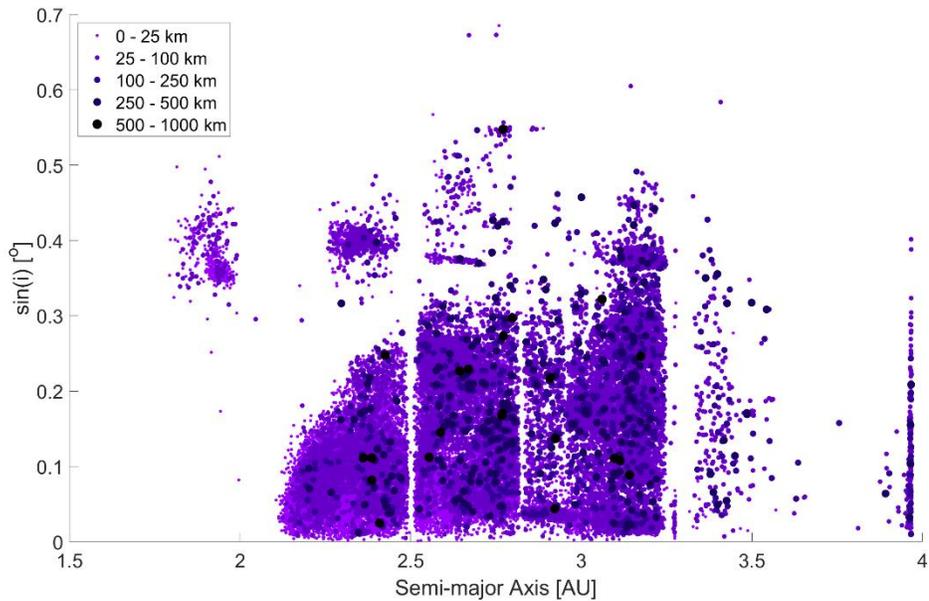

**Figure 1:** Main Belt asteroids synthetic proper elements (first 50,000 numbered bodies; Novakovic, Knezevic and Milani June 2017). Here we have plotted the known asteroid in the Main Belt by semi-major axis and sine(inclination). The size of the point indicates the size of the object.

Our emerging understanding is that planetesimals formed directly from cm-scale pieces into objects of 50-200 km diameter. Meteoritical evidence shows that these planetesimals seem to have formed in at least two distinct reservoirs, with the formation of Jupiter likely to have separated two of these reservoirs [1, 2]. Some of these planetesimals continued to grow, forming planetary embryos (or "protoplanets"), some of which experienced additional accretion to form the terrestrial planets. Planetary migration served to scatter and transport unaccreted planetesimals and planetary embryos, leaving roughly 150-200 in stable orbits between Mars and Jupiter [3]. While subsequent collisions and orbital evolution via gravitational and non-gravitational forces resulted in the millions of objects in the main asteroid belt, these objects are thought to derive from this relatively small number of original planetesimals, some of which remain intact and available for exploration today.

The last decade saw several important advancements in our understanding of Main Belt asteroids and their relevance to the Solar System's history and evolution. We will briefly describe some of these important insights. This is by no means an exhaustive list and are not listed in any particular order.

*Missions: Ceres and Vesta.* The Dawn mission investigated protoplanet Vesta (2011-2012) and dwarf planet Ceres (2015-2018) via orbital encounters that produced near global mapping of these bodies with visible imaging, infrared spectroscopy, elemental spectroscopy, and gravity science [4]. Dawn was the first spacecraft to use solar electric propulsion for a planetary science mission. Key results at Vesta include a better understanding of early planetary differentiation as a consequence of short-lived radioisotope decay. At Ceres, Dawn discovered a relict ocean world and potentially ongoing brine effusion [5, 6]. Spectroscopic and elemental measurements at Vesta confirmed earlier conclusions that Vesta is consistent with the properties expected of the HED parent body [e.g., 7] although the definitive identification of Vesta as "the" HED parent body, or indeed any single asteroid to a group of meteorites, may be an oversimplification. These same techniques at Ceres demonstrated the existence of regolith with latitude-dependent ice content and an interpretation of a CI-like surface composition [8], with the complication of a widespread ammoniated component that has not been found in any meteorite samples [8, 5]. Spectral measurements by Dawn have also detected high local concentrations of organic materials on Ceres [9, 10, 11].

*The Asteroidal-Cometary Continuum:* The neat division of small bodies into the rocky asteroids and the icy comets began to be blurred in the later years of the previous decade, and in this decade evidence accumulated that a large number of objects in the main belt are at least somewhat icy in nature. We discussed Ceres above, and observations of 10 Hygiea by multiple groups [12, 13] have found it to have a spectrum in the 3-µm region very similar to that of Ceres, implying similar hydrated minerals and volatiles and perhaps a similar history and nature. Absorption bands on 24 Themis [14, 15] and 65 Cybele [16] were interpreted as due to water ice frost and organic materials, and several additional objects with similar band shapes have been found in the last decade [17, 12]. Spectra of 324 Bamberga showed greater similarity to comet 67P, the target of the Rosetta Mission, than any other asteroids in the 3-µm region [13]. The spectral evidence for icy asteroids was reinforced by observations of coma on some small asteroids. At least some of these objects (but not all) appeared to have activity driven by sublimation, implying near-surface volatiles and recent exposure.

*Grand Tack Dynamical Model:* Dynamical studies over the past decade have shown that small bodies both are tracers of planetary migration early in solar system history and were drivers of it [e.g., 18]. In the "Grand Tack" scenario, inner-solar-system planetesimals and outer-solar-system planetesimals are transported both into and out of the present-day asteroid belt, with an implication that low-albedo, volatile-rich objects were originally formed among the giant planets (and the relatively rare D-class asteroids were transported from the transneptunian region) while the higher-albedo, volatile-poor objects were originally formed sunward of Jupiter. The NEOWISE survey demonstrated that low-albedo material dominates the asteroid belt [19, 20], though the spectral evidence discussed in the previous section shows that low-albedo objects have diverse compositions.

*Primordial Families:* Identification and interpretation of asteroid dynamical families were subject to great advances during the last decade, thanks in part to both an increase in observational data [e.g., 21] as well as a greater appreciation of the Yarkovsky Force [22, 23]. Recently, multiple authors have proposed that large subsets of the main belt population derive from a very small number of parent bodies: Delbo et al. [3] concluded that practically all low-albedo objects in the inner Main Belt belong to a single "Primordial Family", with a similar family uniting most medium-albedo, X-complex objects in the middle belt [24]. Dermott et al. [24] found that 85% of inner-belt objects can be traced to one of only five families. These findings suggest that a comprehensive understanding of the compositions of the vast majority of main-belt asteroids may be undertaken via investigations of a relatively small number of objects, which based on other work quoted above likely represent original planetesimals.

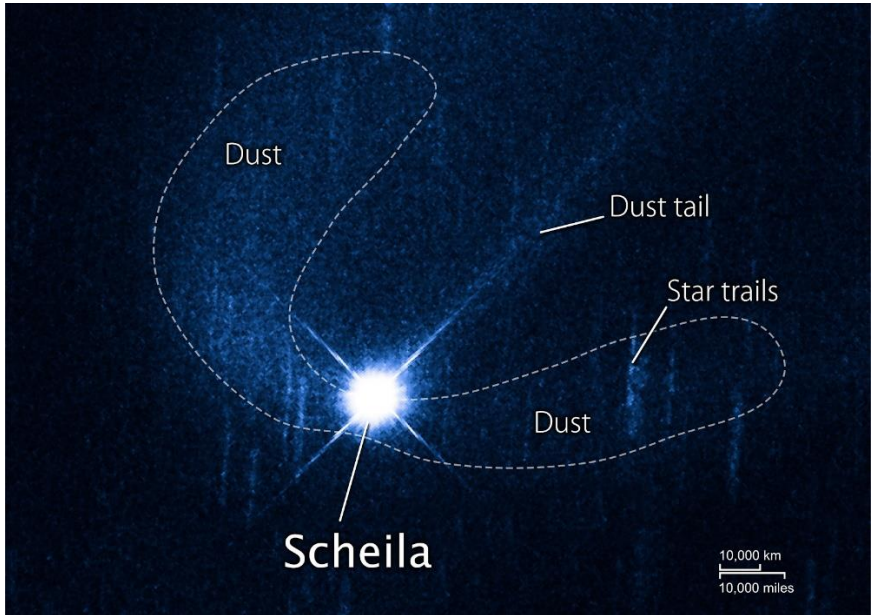

**Figure 2:** Asteroid (596) Scheila with impact driven activity. Image credit: NASA, ESA and Jewett.

*Non-Sublimating Active Asteroids and Their Drivers:* The greater understanding of the Yarkovsky Force comes hand-in-hand with greater understanding of other non-gravitational forces and their influence on asteroid geophysics. The YORP torque has been identified as the likely cause for asteroid satellites and possibly the "top-shapes" that appear to be common in ~km-scale near-Earth objects [e.g., 26]. YORP spin-up and disruption/fission events also appear to be the cause of activity in most of those active asteroids that are not experiencing sublimation. The number of known MBAs is large enough, and their monitoring (via all-sky surveys) frequent enough, that discovery and observation of these events have become relatively routine. Other events appear to be due to impacts, including one on the large asteroid (596) Scheila [e.g., Figure 2; 27]. Particle ejection events on Bennu observed by OSIRIS-REx may also be impact related [e.g., 28], but we do not have appropriate data available from any other object to know whether Bennu's activity is unique or widespread.

Finally, thermally-driven fracturing forces appear to be important on NEOs, possibly driving activity on 3200 Phaethon, and likely causing breakdown of blocks into smaller pieces.

*Looking Forward:* The scientific advances in the last decade belie the wealth of questions that remain unanswered with respect to the Main Asteroid Belt and the evolution of the Solar System that can only be accessed through these bodies. The significant unanswered questions facing the community broadly fall into three areas: physical properties and processes of asteroid surfaces, chemical composition and evolution of asteroids and dynamical evolution of the Main Belt and Solar System. At the broadest level, the following questions require further investigation in the next decade:

- **Fundamental Question 1:** What are the physical properties and key processes (e.g., differentiation, hydrothermal activity, impact cratering, tectonics, regolith development, non-gravitational forces, and space weathering) on asteroids and how are they modified over time?
- **Fundamental Question 2:** What are the current compositions of asteroids, what do these compositions imply about the initial conditions of planet formation and how have asteroid compositions evolved over time?
- **Fundamental Question 3:** What is the distribution of asteroids, today, and how has material migrated from where it initially formed? What was the nature of the main belt after planetary migration ended, and how has collisional evolution created today's size-frequency distribution?

Please see the **Supplementary Table 1** that describes more exact questions within each topic.

*Recommendations:*

**We recommend a balanced program of telescopic observation (ground-based, airborne, and space-based), laboratory studies, theoretical research and missions to Main Belt Asteroids utilizing the full spectral range from ultraviolet to far-infrared to investigate these outstanding fundamental questions.**

*Observational Research:* Observational research, spanning wavelengths from ultraviolet to far-infrared, is a key methodology for studying asteroids, particularly for answering aspects of all of the fundamental questions listed above. Observational research provides a wealth of data on the asteroid population and its characteristics. Facilities of particular importance to asteroids science include: Arecibo (Puerto Rico) and Goldstone Solar System (California) radar telescopes, The Lowell Discovery Telescope (Arizona), the NASA IRTF on Mauna Kea (Hawaii), Adaptive Optics on the W.M Keck Observatory & Gemini Observatory, Pan-STARRS (Hawaii), Catalina Sky Survey , Vera C. Rubin Telescope (Chile), Los Cumbres Observatory (world-wide), the Stratospheric Observatory For Infrared Astronomy (SOFIA) as well as Space-based platforms including the Hubble Space Telescope, James Webb Space Telescope and NEOWISE. We acknowledge that these ground-based facilities are located, or in the case of SOFIA, maintained and operated on land of Indigenous communities and First Peoples[1] including: Hawaii: Kō Hawaiʻi Paeʻāina; Puerto Rico: Boriken Taino; Arizona: Hohokam, Western Apache, Pueblos, Zuni, Hopi and Diné, Sobaipuri, Tohono O'odham - Papago, O'odham; California: Newe (Western Shoshone) Fernandeño Tataviam, Yuhaviatam Maarenga'yam (Serrano), Kizh; New Zealand: Ngai Tahu; Chile: Gününa Küne. Without the land that these facilities occupy and the stewardship of these communities, we could not perform much of the science past or future. Furthermore, while these telescope assets are crucial for the advancement of our science, we recognize that their presence is not always welcomed or sanctioned by the Indigenous communities and First Peoples, particularly in Hawaii. Additionally, the contribution of backyard astronomers through citizen science programs dedicated to coordinate hundreds of small telescopes to observe asteroids by direct imaging or occultation is key to connect the public with our research.

We echo white papers in the Astro2020 Decadal Survey that identified the significance of using NASA Astrophysical facilities for planetary science. JWST is poised to greatly enhance the investigation of large planetesimals, particularly with its spectroscopic capabilities [e.g., 29]. The Wide Field Infrared Survey Telescope, similarly, will provide capabilities to observe small, faint asteroids in the 0.43-2.0-micron region which is difficult to do from the ground [e.g., 30]. We support the following capabilities for upcoming missions [e.g., 31]: a broad range of wavelengths from the UV to the far-IR, non-sidereal tracking, and schedule flexibility to support Main Belt Asteroids science, in particular.

As new asteroids continue to be discovered and observed, the Minor Planet Center and the JPL Center for NEO Studies record, track, and catalog the asteroid population and support planetary defense assessments. These organizations are critically important for observing and understanding Main Belt asteroids.

*Laboratory Studies*: Continued and new support for facilities to characterize and curate meteorites and returned samples from asteroids is essential for answering the fundamental science questions in the upcoming decade [see also 32]. Sustained support for laboratory studies that measure optical constants of minerals and volatiles is key to understanding the composition of asteroids. Furthermore, continued and new support for meteorite searches including the ANSMET program as well as searches in other regions including the Northern Sahara, Saudia Arabian Peninsula, and Atacama Deserts [e.g., 33] are required to find meteorites. A full list of experimental work required for advancing the understanding of small bodies is found in [34].

*Theoretical Research:* Theoretical research represents some of the most important advances in asteroid science in the last decade and will certainly continue to be on the forefront of our field moving forward. These advances are in both dynamical evolution of the Solar System [e.g., 35] and geophysical modeling [e.g., 36. 37, 38] To support theoretical studies related to asteroids we require continued and new support for theoretical studies of dynamics relevant for Main Belt Asteroids. Facilities include computing resources including superclusters and software development for theoretical work.

---

[1] Native Land: https://native-land.ca/ accessed 7/2/2020

*Missions to Main Belt Asteroids:* Missions provide critical, up-close looks at individual asteroids or, in the case of the Lucy Mission, a small number of asteroids. Currently, there is one planned Discovery-class mission to a Main Belt asteroid, (16) Psyche (*Psyche Mission*). This mission will provide a greater understanding of Psyche and asteroids with similar near-infrared spectra throughout the Main Belt. (16) Psyche and the two Dawn mission targets are each unique in the Main Belt. Additional flyby, rendezvous, and sample return, particularly from objects not well-represented in the meteorite population or bodies that may have substantial material formed in other planetary systems (e.g., formed in supernovae, interstellar medium or in other planetary systems; [39]) would provide critical new scientific insights into more recent theoretical and numerical work. Conversely, while NEOs with chondritic compositions have been visited, missions to main-belt objects or families consistent with ordinary or carbonaceous chondritic compositions would provide valuable information about the geology and geophysics of these compositions on planetesimal-sized bodies, and will be representative of the vast number of objects in the asteroid population.

While many aspects of the Fundamental Questions in Main Belt Asteroid science can be answered using means currently available to us now (e.g., extant telescope facilities and modeling), missions to asteroids are critical for true advancement of the field. There are a variety of measurements that missions uniquely make. These include gravity and density particularly for asteroids not in multiple systems (e.g., NEAR at Mathilde), quantitative hydrogen abundance via neutron spectroscopy (important for studying volatiles, especially water), iron-to-silicon ratio, iron-to-oxygen ratio, sulfur, and other elemental abundance measurements via X-ray or gamma-ray spectroscopy, high-resolution imaging to investigate boulder and crater size distributions, geomorphology and geophysics (e.g., crater rays, plumes, topography and stratigraphy), surface composition variations through high spatial-resolution infrared spectroscopy (e.g. Occator crater on Ceres), and mass spectroscopy (similar to the Rosetta Mission) especially for active asteroids.

Discovery-class missions have revolutionized small body science, and continue to be valuable as a cost-effective, rapid-cadence means of addressing high-priority science objectives in planetary science (e.g., NEAR, Dawn, Lucy, and Psyche). We fully support more Discovery Class missions to Main Belt Asteroids because of the rich and diverse science offered by that class of objects that benefits the community at large. Impressive missions to main-belt targets are achievable within the cost caps that have been in place, as demonstrated by Dawn and anticipated with Psyche.

We, further, support the development of New Frontiers level missions to Main Belt Asteroids to investigate the Fundamental Questions described above. Within our diverse community, missions to Main Belt Asteroids are strongly supported[2]. Particularly, missions to return to Ceres and a multiple main belt rendezvous are highly rated within the Small Bodies Assessment Group and Ceres is recognized as a stepping stone for the exploration of ocean worlds in the Roadmap to ocean worlds [40]. We fully support these missions which could answer both

---

[2] SBAG Community Survey (2020) https://www.lpi.usra.edu/decadal/sbag/

the Fundamental Questions outstanding in our field as well as other high priority questions in the planetary science community. Indeed, asteroids in the Main Belt represent pristine materials from the earliest times in the Solar System's history. They are laboratories for studying many aspects of terrestrial planets (e.g., core formation), mid-sized icy bodies (e.g., differentiation), that can be accessed nowhere else. At the boundary between the inner and outer solar system, they are also witnesses of the planetesimal migration events in the early solar system [41]

**Supplementary Table 1:** Traceability Matrix of fundamental and key sub-questions facing Main Belt Asteroid science in the upcoming decade:

| Areas of research | Fundamental question | Key questions | SBAG Objectives[3] |
|---|---|---|---|
| *Physical Processing of Main Belt Asteroids* | 1. What are the physical properties and key processes (e.g., differentiation, hydrothermal activity, impact cratering, tectonics, regolith development, non-gravitational forces, and space weathering) on asteroids and how are they modified over time? | 1a. What are the physical properties of asteroid surfaces and how are asteroid surfaces modified over time? | 1.1.1, 1.2.4, 1.2.5, 1.3.1, 1.3.2, 1.3.3, 1.3.5, 1.4.1, 1.4.2, 1.4.5 |
| | | 1b. How does microgravity/low gravity environments effect physical structure of Main Belt Asteroid surfaces? | 1.3.3, 1.3.4, 1.4.1, 1.4.2, 1.4.5 |
| | | 1c. How do porosity (esp. compaction, fairy castles), grain size and composition effect on surface structure? | 1.3.3, 1.3.4, 1.4.1, 1.4.2, 1.4.5 |
| | | 1d. How do shocks/impacts affect the surfaces of asteroids? | 1.1.1, 1.2.5, 1.3.1, 1.3.2, 1.3.3, 1.3.4, 1.3.5, 1.4.1, 1.4.2, 1.4.5 |
| | | 1e. What is the distribution of physical properties in the Main Belt (e.g., grain size, surface condition, shocked materials, boulders, etc.)? | 1.2.1, 1.2.2, 1.2.3, 1.2.5, 1.3.3, 1.3.5, 1.4.1, 1.4.2 |
| | | 1f. What are the granular mechanics of Main Belt Asteroid surfaces? | 1.2.5, 1.3.3, 1.4.1, 1.4.2, 1.4.5 |
| | | 1g. What can we learn about asteroidal interiors from remote sensing of their surfaces? What are the geophysical expressions of large-scale processing (e.g., | 1.2.5, 1.3.1, 1.3.3, 1.4.3, 1.4.4 |

---

[3] 2020 SBAG Goals Document: https://www.lpi.usra.edu/sbag/goals/

| | | aqueous alteration, thermal metamorphism, differentiation)? | |
|---|---|---|---|
| | | 1h. How does size drive internal evolution in D>100 km bodies and compositional fractionation? Did large planetesimals (or still do in the case of Ceres) host long-lived mudball-type convection? | 1.2.4, 1.2.5, 1.3.5, 1.4.1, 1.4.2, 1.4.3 |
| | | 1i. How are materials or masses formed in asteroidal interiors changed by the process that bring them to asteroidal surfaces? | 1.3.3, 1.4.1, 1.4.2 |
| *Geochemical processing of Main Belt Asteroids* | What are the current compositions of asteroids, what do these compositions imply about the initial conditions of planet formation and how have asteroid compositions evolved over time? | 2a. What was the compositional gradient of asteroid (and asteroid parent body) formation locations during initial protoplanetary accretion, and what was the redox and thermal state/gradient of the early Solar System? | 1.1.1, 1.1.2, 1.1.3,1.1.4, 1.2.3, 1.2.4, 1.2.6, 1.3.1, 1.3.2, 1.3.3, 1.4.2, 1.4.3 |
| | | 2b How did the initial compositional gradient affect planetary formation and evolution? | 1.1.3, 1.1.4, 1.2.2, 1.2.3, 1.2.4, 1.2.6, 1.4.3, |
| | | 2c. What was the distribution of volatiles in the early Solar System, and what role did asteroids play in the delivery of water and organics to the inner Solar System? | 1.1.3, 1.1.4, 1.2.3, 1.2.4, 1.2.6, 1.3.1, 1.3.2, 1.3.3, 1.3.4, 1.3.5, 1.4.2, 1.4.3, |
| | | 2d. What are the key processes (e.g., differentiation, hydrothermal activity, impact cratering, tectonics, regolith development, non-gravitational forces, and space weathering) on | 1.1.1, 1.1.3, 1.2.4, 1.2.5, 1.3.1, 1.3.2, 1.3.3, 1.3.4, 1.3.5, 1.4.1, 1.4.2, 1.4.3, 1.4.4, 1.4.5 |
| | | asteroids and how are they modified over time? How do these processes affect the surfaces and interiors of asteroids? | |
| | | 2e. Are the surfaces (especially the top 10s of microns observable remotely via optical, infrared, or X-ray techniques) compositionally representative of the whole asteroid? | 1.1.1, 1.1.2, 1.1.4, 1.2.4, 1.3.1, 1.3.2, 1.3.3, 1.3.4, 1.3.5, 1.4.1, 1.4.2, 1.4.5 |
| | | 2f. What is the chemical/mineralogical changes caused by aqueous alteration, thermal metamorphism, differentiation on Main Belt Asteroids? How do these processes affect both the small and large scale? How do these processes affect the surfaces and interiors of asteroids? | 1.1.1, 1.1.2, 1.2.4, 1.3.1, 1.3.2, 1.3.3, 1.3.5, 1.4.3, 1.4.4 |
| | | 2g. What is the geophysical expression of large-scale processing (e.g., aqueous alteration, thermal metamorphism, differentiation)? | 1.2.5, 1.3.1, 1.3.3, 1.3.5, 1.4.3, 1.4.4 |
| | | 2h. How do shocks/impacts affect chemical composition of Main Belt Asteroid surfaces? | 1.1.1, 1.2.4, 1.2.5, 1.3.1, 1.3.2, 1.3.3, 1.3.4, 1.3.5, 1.4.1, 1.4.2 |
| | | 2i. What are the characteristics of water-rich and/or hydrated asteroids and how have the volatiles on those asteroids evolved? | 1.1.1, 1.1.2, 1.1.3, 1.2.4, 1.2.5, 1.3.1, 1.3.2, 1.3.3, 1.3.4, 1.3.5, 1.4.3 |
| | | 2j What are the effects of space weathering on composition and physical presentation of asteroid surfaces? | 1.1.1, 1.1.2, 1.1.3, 1.1.4, 1.2.4, 1.2.5, 1.3.1, 1.3.2, 1.3.3, 1.3.5, 1.4.1, 1.4.2 |
| | | 2k. What are the bulk densities, porosities, and moments of inertia of Main Belt asteroids? How are these properties related to primary parent body processes (e.g., thermal metamorphism, aqueous alteration, differentiation, core formation, etc.)? | 1.4.1, 1.4.2, 1.4.3, 1.2.5 |
| | | 2l. What is the distribution of chemical properties of asteroids in the Main Belt? | 1.4.1, 1.4.2, 1.2.5, |
| *Dynamical Processing of Main Belt Asteroids* | What is the distribution of asteroids, both near-Earth and Main Belt, today, and how has material migrated from where it initially formed? What was the nature of the main belt after planetary migration ended, and how has collisional evolution created today's size-frequency distribution? | 3a. What was the compositional gradient of asteroid formation locations during initial protoplanetary accretion, and what was the redox and thermal state/gradient of the early Solar System? How did this affect planetary formation and evolution? | 1.4.1, 1.4.2, 1.2.5, |
| | | 3b. What was the nature of the main belt after planetary migration ended, and how has collisional evolution created today's size frequency distribution? | 1.4.1, 1.4.2, 1.2.5, |
| | | 3c. What was the distribution of volatiles in the early Solar System, and what role did asteroids play in the delivery of water and organics to the inner Solar System? | 1.4.1, 1.4.2, 1.2.5, |
| | | 3d. How are meteorites delivered to Earth? Are all populations of asteroids represented in this record? | 1.4.1, 1.4.2, 1.2.5, |
| | | 3e. What are the dynamical processes that affect asteroid families? | 1.4.1, 1.4.2, 1.2.5, |
| | | 3f. What dynamical processes affected the early Solar System's history? What is the dynamical | 1.4.1, 1.4.2, 1.2.5 |
| | | evolution of the Solar System? | |
| | | 3g. How has material migrated from where it initially formed? | 1.4.1, 1.2.5 | 1.4.2, |
| | | 3h. What can we generalize about the Main Asteroid Belt's role in planet formation to other solar systems? | 1.4.1, 1.2.5, | 1.4.2, |
| | | 3i. What is the relationship between main-belt asteroids and other small-body populations in the solar system? | 1.2.2, 1.2.4, | 1.2.3, |